\documentclass[12pt]{article}
\usepackage{graphicx}
\usepackage{subfigure}
\usepackage{amsmath,amsthm}
\usepackage{amssymb,amsfonts}
\usepackage{authblk}
\usepackage{url}
\usepackage{booktabs}

\newtheorem{definition}{\textbf{Definition}}

\newtheorem{theorem}{\textbf{Theorem}}

\newtheorem{corollary}{\textbf{Corollary}}

\title{Associations between finger tapping, gait and fall risk with application to fall risk assessment}
\author{Jian MA\thanks{Email: majian@hitachi.cn}}
\affil{Hitachi (China) Research \& Development Corporation}

\begin{document}

\maketitle

\begin{abstract}
\noindent
As the world ages, elderly care becomes a big concern of the society. To address the elderly's issues on dementia and fall risk, we have investigated smart cognitive and fall risk assessment with machine learning methodology based on the data collected from finger tapping test and Timed Up and Go (TUG) test. Meanwhile, we have discovered the associations between cognition and finger motion from finger tapping data and the association between fall risk and gait characteristics from TUG data. In this paper, we jointly analyze the finger tapping and gait characteristics data with copula entropy. We find that the associations between certain finger tapping characteristics ('number of taps', 'average interval of tapping', 'frequency of tapping' of both hands of bimanual inphase and those of left hand of bimanual untiphase) and TUG score are relatively high. According to this finding, we propose to utilize this associations to improve the predictive models of automatic fall risk assessment we developed previously. Experimental results show that using the characteristics of both finger tapping and gait as inputs of the predictive models of predicting TUG score can considerably improve the prediction performance in terms of MAE compared with using only one type of characteristics.
\end{abstract}


\section{Introduction}
As the world is continuously aging \cite{un2017}, elderly care becomes a high concern of the society. Delivering better care to the elderly can not only improve their wellbeings of all aspects, but also relieve the burden of their family and society. At the core of all the needs of the elderly is the need for health and medical care. According to the WHO study \cite{who2013}, dementia and fall injury are two main diseases which suffer the elderly mainly instead of other populations. How to manage these diseases for the elderly is a main challenge. To address these issues, better care instruments are very needed. 


In our previous research, two technologies for smart elderly care were developed: one for predicting dementia \cite{ma2019} and the other for automatic fall risk assessment \cite{ma2020}. The former technology is to predict Minimal Mental State Examination (MMSE) score from finger tapping measurement with machine learning, in which a group of characteristics of finger tapping movement are extracted and selected for the predictive models. The latter technology is based on the similar machine learning methodology, to predict Timed Up and Go (TUG) score from a group of gait characteristics extracted from video with steoro vision and 3D pose estimation technologies.

In these research, finger tapping test and TUG test are two sources of experimental data. The two types of data were analyzed separately in two independent works. In fact, cognitive impairment and fall are both common syndromes of the aging people. There are many evidences that cognition impairment is an predictor of fall risk and associated with increased fall risk \cite{muir2012}. There are also several research reporting the relationship between cognition and gait \cite{odasso2012,valkanova2017,peel2019}. 

In this paper, we will continue to investigate the relationship between finger tapping, cognition, gait and fall risk by jointly analyzing the data collected from the two research in \cite{ma2019,ma2020}. Previously, the relationship between characteristics of finger tapping and cognition impairment has been studied, in which certain characteristics (number of taps, average interval of tapping, frequency of tapping, and SD of average interval of tapping) were found to be associated with MMSE score \cite{ma2019}. However, the other relationships between cognition, finger motion and gait remain to be studied. 

Discovering such associations is of fundamental importance for automatic fall risk assessment because they can lay scientific foundations of the predictive models built in the research. If, for example, the relationship between finger tapping and TUG score is found, then the models for predicting TUG score can be improved with characteristics of finger tapping. Such relationship will also be the evidence that finger motor ability and functional ability are related with each other. In this paper, we will try to find such associations within our dataset and then utilize the association relationships to improve the prediction of fall risk, i.e., TUG score.

The mathematical tool used in this research is Copula Entropy (CE), which is defined by Ma and Sun \cite{ma2008}. It is a rigorously mathematical concept for statistical independence testing and enjoys several good axiomatic properties for statistical independence measure. A simple non-parametric method for estimating CE was also proposed, which makes CE universally applicable without making any assumptions \cite{ma2008}. As a tool for discovering association relationships, it has been applied successfully in our previous research to study the relationship between finger motor and cognitive ability \cite{ma2019} and the relationship between gait characteristics and fall risk \cite{ma2020}.

The main contributions of this paper include:
\begin{itemize}
\item The association relationships between finger tapping, gait, and fall risk are discovered with CE. Particularly, the association between certain gait characteristics and MMSE, and the association between certain characteristics of finger tapping and fall risk are discovered;
\item A method for predicting TUG score (fall risk) with both the characteristics of finger tapping and gait is proposed and its advantage is demonstrated on real data. The predictive models such developed are explainable due to the association relationships discovered above.
\end{itemize}

This paper is orgnized as follows: in Section \ref{s:CopEnt}, the theory and estimation of CE will be introduced; Section \ref{s:data} will give the details on the data collected from the previous research; experiments and results on association discovery and its application on automatic TUG score prediction will be presented in Section \ref{s:exp} and followed by some discussion in Section \ref{s:dis}; finally, we conclude the paper in Section \ref{s:con}.

\section{Copula Entropy}
\label{s:CopEnt}
\subsection{Theory}
\noindent
Copula theory unifies representation of multivariate dependence with copula function \cite{nelsen2007,joe2014}. According to Sklar theorem \cite{sklar1959}, multivariate density function can be represented as a product of its marginals and copula density function which represents dependence structure among random variables. This section is to define an association measure with copula. For clarity, please refer to \cite{ma2008} for notations.

With copula density, Copula Entropy is define as follows \cite{ma2008}:
\begin{definition}[Copula Entropy]
\label{d:ce}
	Let $\mathbf{X}$ be random variables with marginals $\mathbf{u}$ and copula density $c(\mathbf{u})$. CE of $\mathbf{X}$ is defined as
	\begin{equation}
	H_c(\mathbf{X})=-\int_{\mathbf{u}}{c(\mathbf{u})\log{c(\mathbf{u})}}d\mathbf{u}.
	\label{eq:ce}
	\end{equation}
\end{definition}

In information theory, Mutual Information (MI) and entropy are two different concepts \cite{infobook}. In \cite{ma2008}, Ma and Sun proved that MI is actually a kind of entropy, negative CE, stated as follows: 
\begin{theorem}
\label{thm1}
	MI of random variables is equivalent to negative CE:
	\begin{equation}
	I(\mathbf{X})=-H_c(\mathbf{X}).
	\end{equation}
\end{theorem}

Theorem \ref{thm1} has simple proof \cite{ma2008} and an instant corollary (Corollary \ref{c:ce}) on the relationship between information containing in joint probability density function, marginals and copula density.
\begin{corollary}
\label{c:ce}
	\begin{equation}
		H(\mathbf{X})=\sum_{i}{H(X_i)}+H_c(\mathbf{X})
	\end{equation}
\end{corollary}
The above results cast insight into the relationship between entropy, MI, and copula through CE, and therefore build a bridge between information theory and copula theory. CE itself provides a theoretical concept of statistical independence measure.

\subsection{Estimation}
\label{s:est}
\noindent
It is widely considered that estimating MI is notoriously difficult. Under the blessing of Theorem \ref{thm1}, Ma and Sun \cite{ma2008} proposed a non-parametric method for estimating CE (MI) from data which composes of only two steps: \footnote{The R package \textbf{copent} for estimating copula entropy is available on the CRAN and also on GitHub at: \url{https://github.com/majianthu/copent}.}
\begin{enumerate}
	\item Estimating Empirical Copula Density (ECD);
	\item Estimating CE.
\end{enumerate}

For Step 1, if given data samples $\{\mathbf{x}_1,\ldots,\mathbf{x}_T\}$ i.i.d. generated from random variables $\mathbf{X}=\{x_1,\ldots,x_N\}^T$, one can easily estimate ECD as follows:
\begin{equation}
F_i(x_i)=\frac{1}{T}\sum_{t=1}^{T}{\chi(\mathbf{x}_{t}^{i}\leq x_i)},
\end{equation}
where $i=1,\ldots,N$ and $\chi$ represents for indicator function. Let $\mathbf{u}=[F_1,\ldots,F_N]$, and then one can derives a new samples set $\{\mathbf{u}_1,\ldots,\mathbf{u}_T\}$ as data from ECD $c(\mathbf{u})$. 

Once ECD is estimated, Step 2 is essentially a problem of entropy estimation which can be tackled by many existing methods. Among those methods, the kNN method \cite{kraskov2004} was suggested in \cite{ma2008}, which leads to a non-parametric way of estimating CE.

\subsection{CE as association measure}
\noindent
Rigorously defined, CE has several properties which an ideal statistical independence measure should have, including multivariate, symmetric, non-negative (0 iff independent), invariant to monotonic transformations, and equivalent to correlation coefficient in Gaussian cases.

Theoretically, CE has many advantages over traditional association measure -- Correlation Coefficient (CC). Implied by definition, CC is a bivariate measure with Gaussian assumption while CE has no such limitations. More theoretical comparisons between CC and CE are listed in Table \ref{t:t1a}. Since CE shows clear advantages over CC, it has been proposed as a method for discovering association relationships \cite{ma2019a}.

\begin{table}
\setlength{\abovecaptionskip}{0pt}
\setlength{\belowcaptionskip}{5pt}
\centering
\caption{Comparisons between CC and CE.}
\begin{tabular}{l|c|c}
	\toprule
	&CC&CE\\
	\midrule
	Linearity&Linear&linear/Non-linear\\
	Order&Second&All\\
	Assumption&Gaussian&None\\
	Dimensions&bivariate&mutlivariate\\
	Association Type&correlation&dependence\\
	\bottomrule
\end{tabular}
\label{t:t1a}
\end{table}

\section{Data}
\label{s:data}
The data used in this paper were collected from 40 subjects recruited at Tianjin, whose age range at 45-84. All the participants signed informed consent. All subjects were administrated to perform four types of test, including Tinetti POMA test, MMSE test, TUG test, and finger tapping test, twice a day for several times in one month. 

The finger tapping test is based on the finger tapping device \cite{kandori2004} which measures finger motion movement with magnetic sensing technique. In each test, four modes of movement are measured: bimanual in-phase, bimanual unti-phase, left hand single, right hand single. For bimanual in-phase and bimanual unti-phase movement, 84 attributes are derived, and for single hand movement, only 40 attributes are derived, as described in \cite{suzumura2016}. In the experiments, each movement lasts for 15 seconds. The characteristics of finger tapping test used in the following experiments include number of taps, average interval of tapping, frequency of tapping and SD of interval of tapping of both hands of bi-inphase and bi-untiphase tapping, which lead to 16 characteristics most associated with MMSE score \cite{ma2019}.

The data on gait include 18 characteristics (mean and SD of the 9 characteristics listed in Table \ref{t:gait}) extracted from video data with the method proposed in \cite{li2019embc}. In detail, the video was recorded during TUG test and then 3D pose was derived by combining 2D pose estimated from video and 3D depth information from 3D cameras. The gait characteristics were calculated from 3D pose series of the whole video of each test.

\begin{table}
\setlength{\belowcaptionskip}{5pt}
	\centering
	\caption{Gait characteristics extracted from video \cite{li2019embc}.}
	\begin{tabular}{l|p{0.5\textwidth}}
		\toprule
		Name&Definition\\
		\midrule
		Gait speed&Speed of body movement\\
		
		Speed variability&standard deviation of stride speeds\\
		
		Stride time&time between one peak and the second-next peak\\
		
		Stride time variability&standard deviation of stride times\\
		
		Stride frequency&median of modal frequency for the ML and half the modal frequencies for the V and AP directions \\
		
		Movement intensity&standard deviation of acceleration rate\\
		
		Low-frequency percentage&Summed power up to a threshold frequency divided by total power\\
		
		Acceleration range&Difference between minimum and maximum acceleration\\
		
		Step length (Pace)&Length of one step\\
		\bottomrule
	\end{tabular}
	\label{t:gait}
\end{table}

After excluding the subjects who did not complete all the four tests, we got 38 subjects and 134 tests totally. Each sample generated from these tests composes of the scores of four tests, the characteristics of finger tapping and gait. In summary, the data includes 134 samples with 4 scores and 34 characteristics.

\section{Experiments and Results}
\label{s:exp}
\subsection{Experiments}
To study the relationship between finger tapping, gait, and fall risk, we conduct an experiment to measure the associations between four scores, the characteristics of finger tapping and gait with CE from the above data. CE \cite{ma2008} is used in this research to measure the associations between characteristics of finger tapping and gait. CE is an ideal tool for measuring statistical dependence in this problem. It is estimated with non-parametric two step method proposed in Section \ref{s:est}. The associations are identified based on the association strength measured by CE.

As its application, we use the identified associations to improve the works on automatic TUG test \cite{ma2020}. We conduct an experiment to study whether the characteristics of finger tapping can be used to improve the models of predicting TUG score that we build in \cite{ma2020}. The characteristics of finger tapping that are most associated with TUG score will be integrated into the predictive models to predict TUG score. 
As contrast, we also conduct an experiment to study the performance of the predictive models for predicting TUG score with only certain characteristics of finger tapping as input. Comparisons on the predictive models with three groups of inputs (finger tapping only, gait only, and both) will be done to check whether integrating the characteristics of finger tapping and gait together can improve the performance of the predictive models. 

In the prediction experiments, only the characteristics which are mostly associated with TUG score are considered, including `number of taps', 'average interval of tapping', 'frequency of taps' of both hands of bimanual inphase from finger tapping test and 4 gait characteristics (including gait speed, pace, speed variance, acceleration range) from the TUG test which has been identified in the previous research \cite{ma2019,ma2020}. 

The predictive models in the experiments are Linear Regression (LR) and Support Vector Regression (SVR) \cite{smola2004}. The ratio between training data and test data are (80/20)\% and the data set was randomly divided for 100 times. The hyper-parameters of SVR are tuned to obtain the best possible prediction results. The performance of the predictive models are measured by Mean Absolute Error (MAE) between the true TUG scores and the predicted scores.

Due to the imbalance deficiency of the MMSE scores in the current data, we can not study whether the characteristics of gait can be used to improve the models of predicting MMSE score that we built in the previous work \cite{ma2019}.

\subsection{Results}
\begin{figure}
	\centering
	\includegraphics[width=0.9\textwidth]{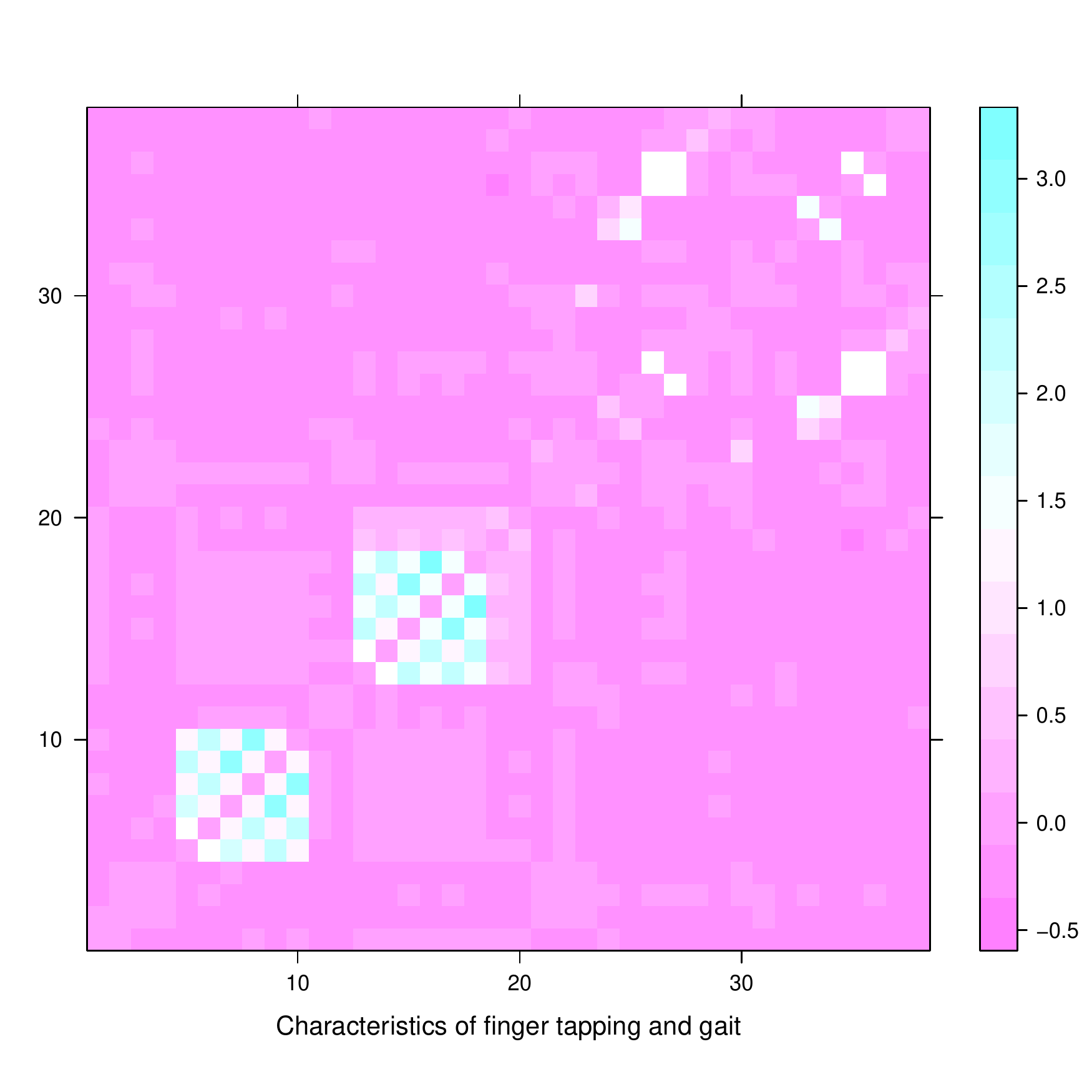}
	\caption{Associations between four scores, 16 characteristics of finger tapping and 18 gait characteristics measured by CE.}
	\label{f:fttugmi1}
\end{figure}

The associations between 4 scores and characteristics of finger tapping and gait measured by CE are shown in Figure \ref{f:fttugmi1}. It can be learned from the Figure that 'number of taps', 'average interval of tapping', and 'frequency of tapping' of both hand of bimanual inphase or bimanual untiphase are associated with each other, which confirmed our finding in the previous research \cite{ma2019}.

\begin{figure}
	\centering
	\includegraphics[width=0.9\textwidth]{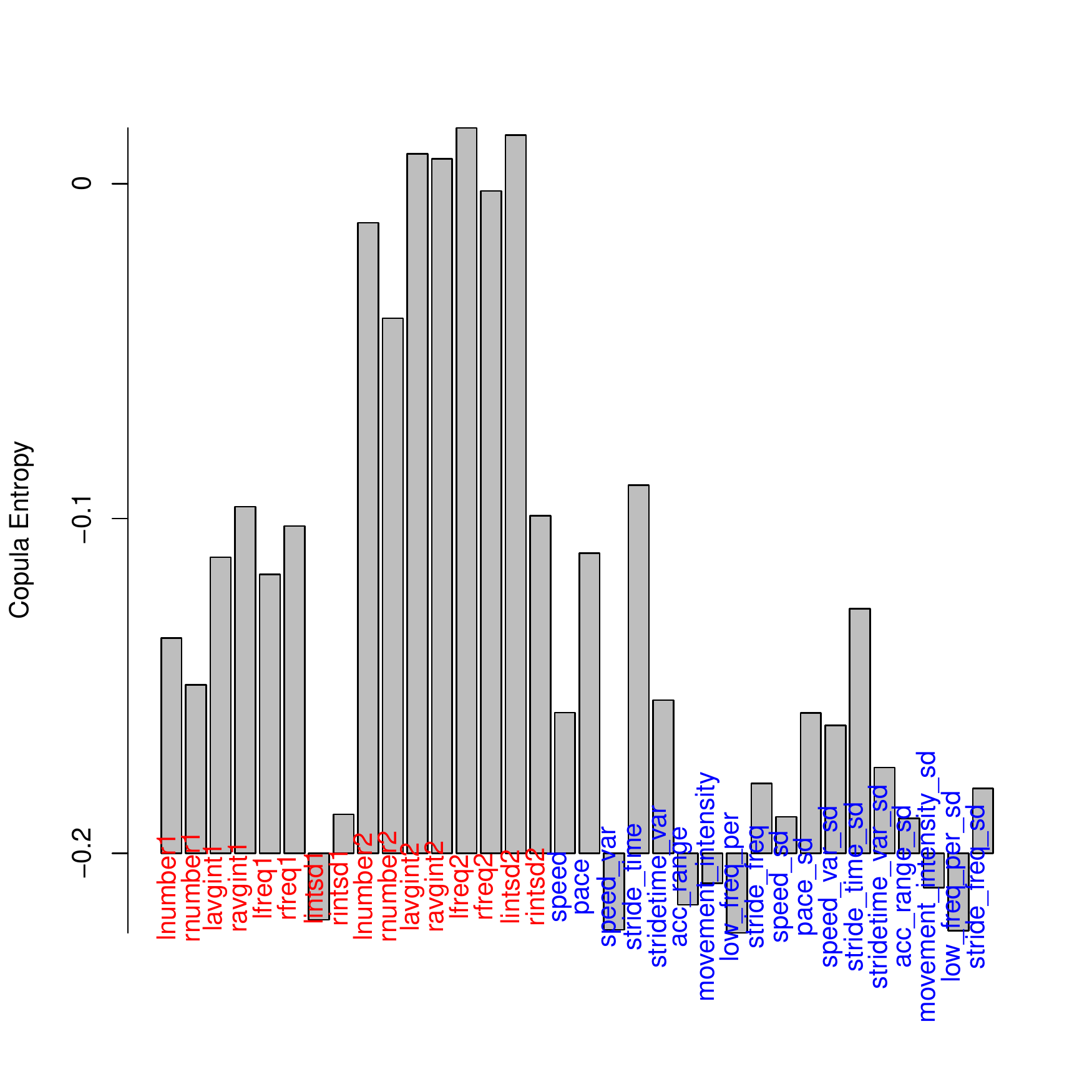}
	\caption{Associations of MMSE score with 16 characteristics of finger tapping and 18 gait characteristics.}
	\label{f:mmsemi1}
\end{figure}

The associations between MMSE score and all the characteristics are shown in Figure \ref{f:mmsemi1}. It can be learned from the Figure that certain characteristics, such as pace and stride time, are also associated with MMSE score with relatively strong strength.

\begin{figure}
	\centering
	\includegraphics[width=0.9\textwidth]{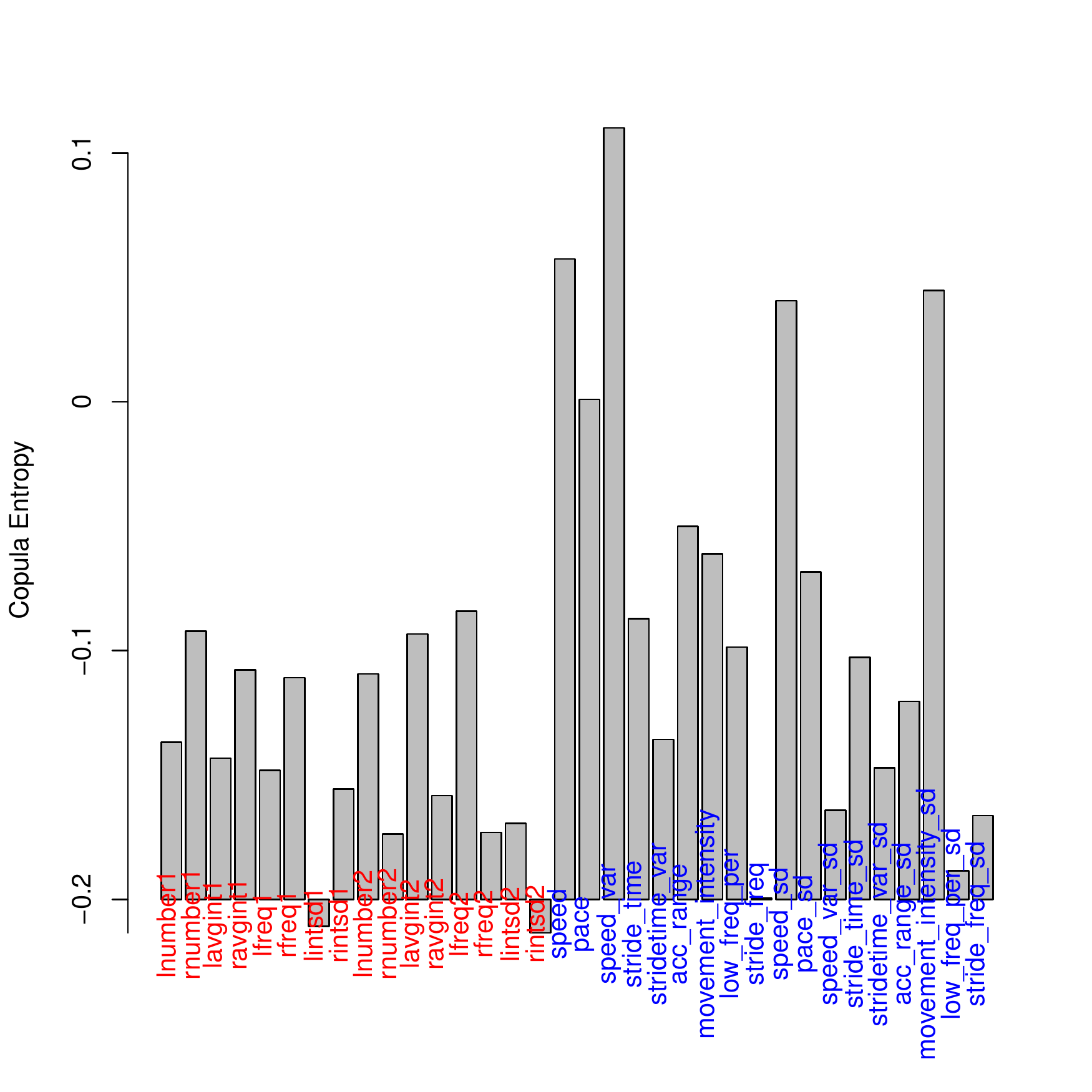}
	\caption{Associations of TUG score with 16 characteristics of finger tapping and 18 gait characteristics.}
	\label{f:tugmi1}
\end{figure}

The associations between TUG score and all the characteristics are shown in Figure \ref{f:tugmi1}. It can be learned from it that 'number of taps', 'average interval of tapping', 'frequency of tapping' of both hands of bimanual inphase and those of left hand of bimanual untiphase are associated with TUG score compared with gait characteristics. The associations between TUG score and characteristics of finger tapping can be interpreted as the associations between finger mobility and functional ability. To the best of our knowledge, this is the first study reporting such relationships. This interesting finding inspires us to use those characteristics of finger tapping to improve the models for predicting TUG score with gait characteristics. 

\begin{figure}
	\centering
	\includegraphics[width=0.9\textwidth]{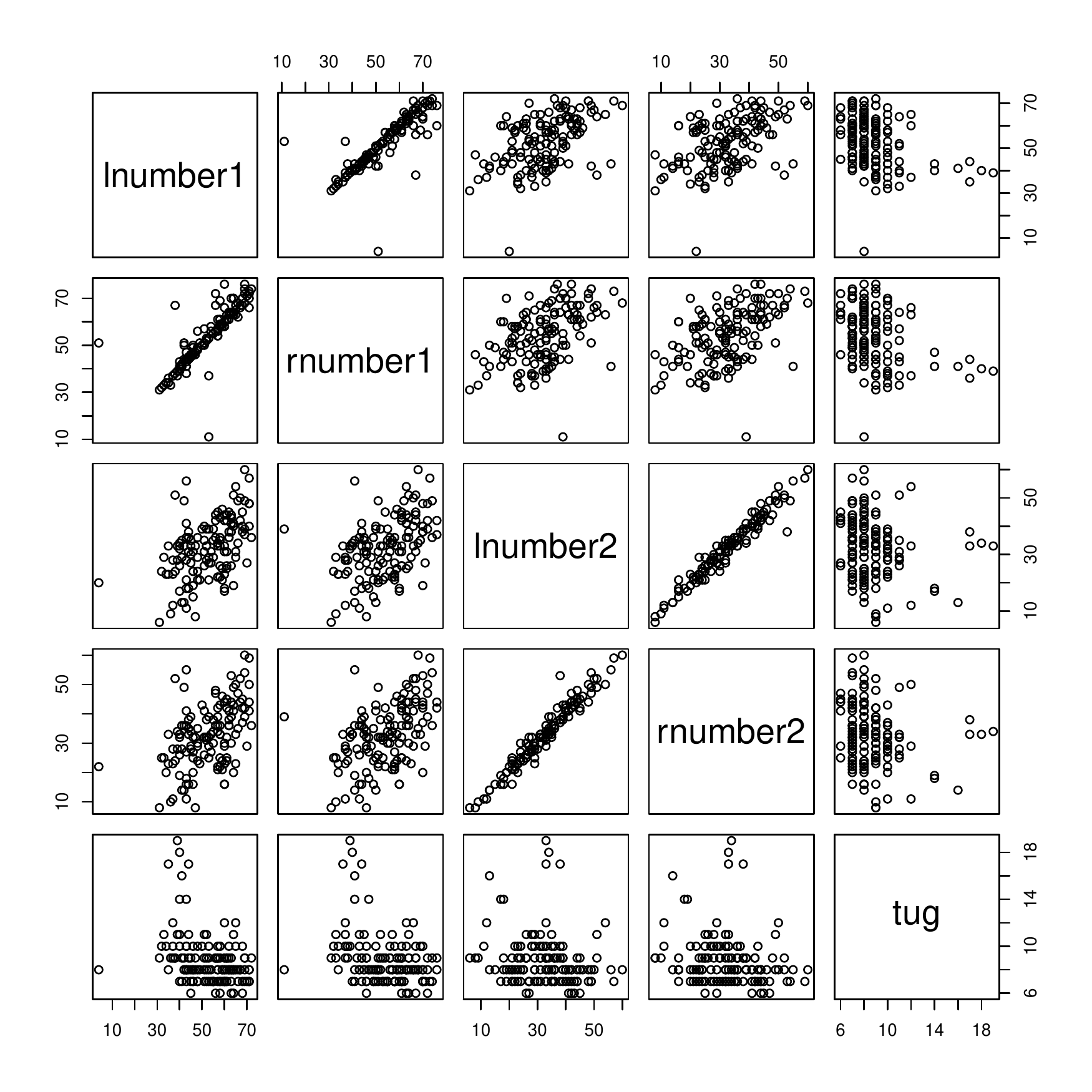}
	\caption{Joint distribution between TUG score and four characteristics of finger tapping (`number of taps' of both hands of bi-inphase and bi-untiphase).}
	\label{f:jointplot1}
\end{figure}

The joint distribution of TUG score and number of taps is shown in Figure \ref{f:jointplot1}, from which it can be learned that the associations between them are highly nonlinear. This suggests that CE, as a nonlinear measure, is the right choice for measuring such associations. 

\begin{figure}
	\centering
	\subfigure[LR]{\label{f:lr1}\includegraphics[width=0.6\textwidth]{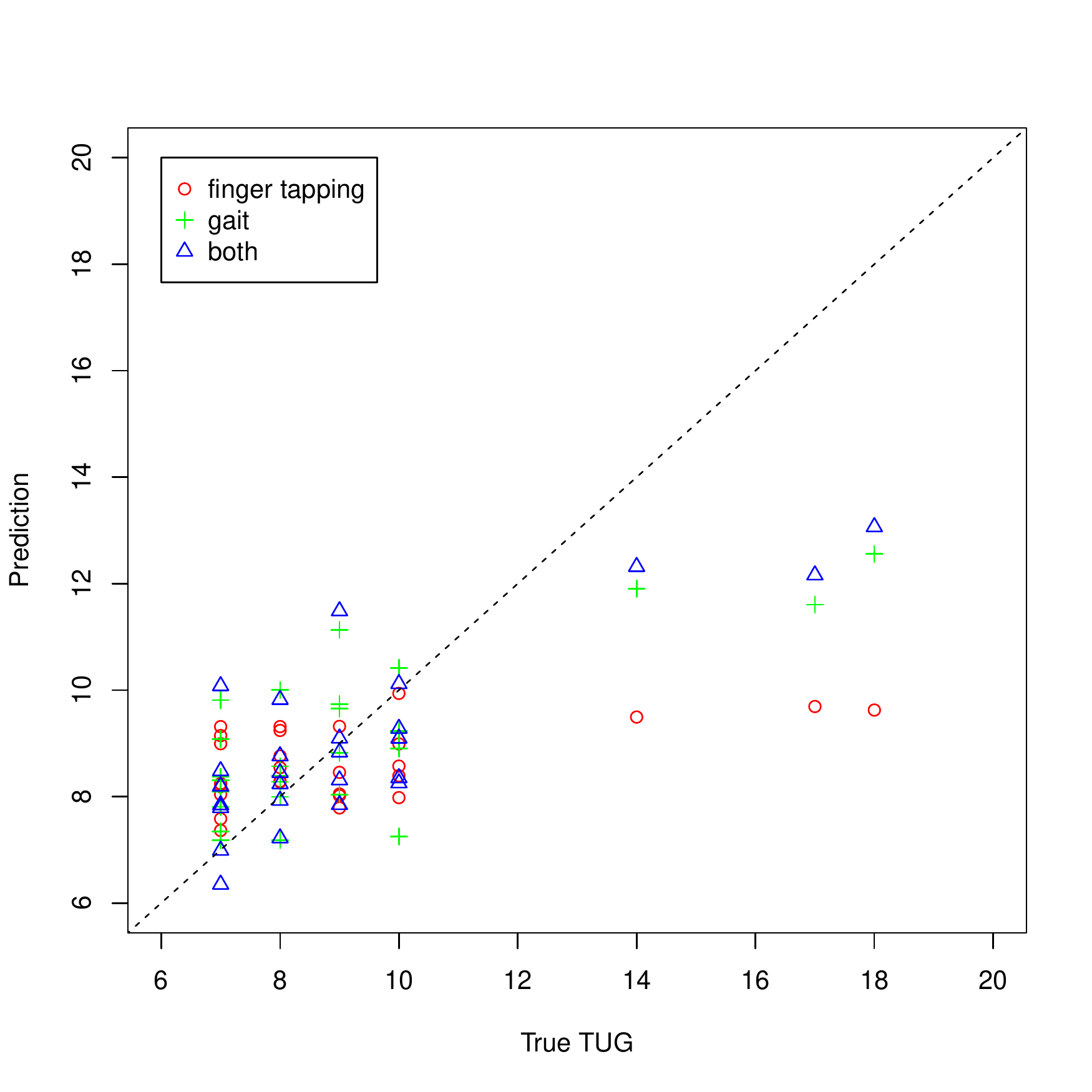}}	
	\subfigure[SVR]{\label{f:svr1}\includegraphics[width=0.6\textwidth]{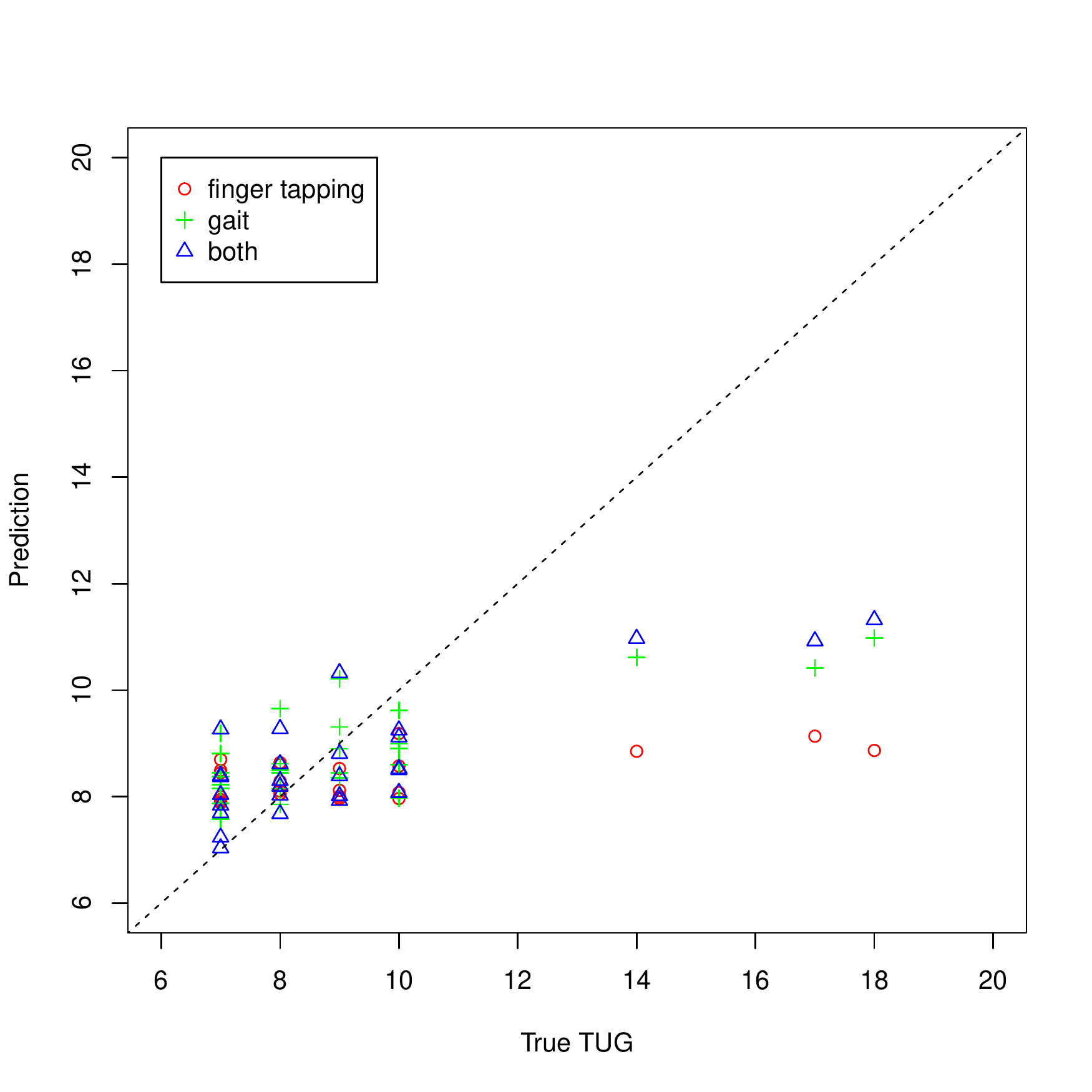}}

	\caption{Performance of the predictive models (LR and SVR) with 3 groups of characteristics as inputs.}
	\label{f:ww}
\end{figure}

We conducted 6 experiments to predict TUG score with different combinations of two models (LR and SVR) and three groups of characteristics (number of taps only, gait only, and both). The prediction results of both models are shown in Figure \ref{f:lr1} and \ref{f:svr1}. It can be learned from both Figures that the prediction with both types of characteristics is considerable better than with only one. We measured the performance of 6 experiments in terms of MAE, as listed in Table \ref{t:perf}. It can be learned from it that: 1) SVR with both characteristics presents the best result with the smallest MAE (=1.218); 2) the performance of SVR is better than that of LR on all the three groups of characteristics; 3) gait characteristics presents stable performance on both LR and SVR while LR presents low prediction performance in terms of MAE with characteristics of finger tapping.

\begin{table}
	\centering
	\caption{Comparison on prediction performance (MAE) between finger tapping, gait or both as inputs of the models.}
	\begin{tabular}{l|cc}
		\toprule
		&LR&SVR\\
		\midrule
		Finger tapping&10.185&1.433\\
		Gait&1.352&1.246\\
		Both&5.560&\textbf{1.218}\\
		\bottomrule
	\end{tabular}
	\label{t:perf}
\end{table}

\section{Discussion}
\label{s:dis}
Figure \ref{f:tugmi1} shows that certain characteristics of finger tapping ('number of taps', 'average interval of tapping', 'frequency of tapping' of both hands of bimanual inphase, and those of left hands of bimanual untiphase are associated with TUG score, which suggests finger motor ability and gait ability are related with each other. To the best of our knowledge, there is no previous research reporting this relationship. Nagasaki et al examined walking patterns and rhythmic movement of fingers of older adults \cite{nagasaki1996}. Hausdorff et al found that even walking and tapping are both automated, rhythmic motor task, the former is shown surprisingly related to catching rather than the latter \cite{hausdorff2005}, which provides an opinion opposite to ours. In Figure \ref{f:tugmi1}, number of taps is also shown strong association with TUG score, which indicate that number of taps is a predictor for fall risk. So it is reasonable to hypothesis that finger motor ability can help to predict fall risk, as this paper does.

Figure \ref{f:mmsemi1} shows that MMSE is also associated with the gait characteristics, such as pace, speed variance and stride time, which implies the relationship between gait and cognition. This result provides another evidence that gait disorder is related to cognitive impairment \cite{odasso2012,valkanova2017,peel2019}. Beauchet et al \cite{olivier2008} reported stride variability is more specific and sensitive in subjects with dementia, which gain moderate support from our experimental results. 

In our experiments, the performance of the SVR model is improved by combining the two characteristics together as inputs of the model. Additionally, the performance of the two models in terms of MAE are different, particularly SVR presents better results than LR does. This may be because that the relationship between the characteristics and TUG score are nonlinear and hence SVR with nonlinear models can perform well. SVR with both types of characteristics presents the best result which suggest that both characteristics are informative for this prediction task. Figure \ref{f:jointplot1} supports this claim.

Remember that the data used in the experiments are unbalanced with a few samples from patients with high fall risk. This may lead to the models tending to predict more positive results rather than fall risk and hence make the MAE unreliable to some extent. Therefore, the performance of the models should be improved with more patients data in the future.

\section{Conclusion}
\label{s:con}
In this paper, we jointly analyze with CE the finger tapping data and gait characteristics data collected from our previous research. The associations between certain finger tapping characteristics, gait characteristics, and TUG score are discovered. These associations are then applied to automatic fall risk assessment to improve the performance of the predictive models on TUG score prediction. Experimental results show that integrating the finger tapping characteristics into these predictive models can considerably improve the prediction performance of the SVR model in terms of MAE. Such associations is an evidence on the relationship between gait, finger motor ability, and functional ability, which lay the scientific foundation of the predictive models.

\section*{Acknowledgements}
The author thanks Zhang Pan and Lin Xiaolie for providing data.

\end{document}